\title{SecureRouting}

\documentclass[journal]{IEEEtran}
\ifCLASSINFOpdf
\else
\fi

\usepackage{times,epsfig,algorithmic,array}
\usepackage{mdwmath}
\usepackage{mdwtab}
\usepackage{eqparbox}
\usepackage{cite}
\usepackage{graphicx}
\usepackage{amsfonts}
\usepackage{epstopdf}
\usepackage{amsfonts}
\usepackage{amsmath}
\usepackage{amssymb}
\usepackage{amsthm}
\usepackage{algorithm}
\usepackage{algorithmic}
\RequirePackage[colorlinks,citecolor=blue,urlcolor=blue]{hyperref}

\begin{document}
%
\title{Secure Location-Aided Routing Protocols With Wi-Fi Direct For Vehicular Ad Hoc Networks}
%
%
%

\author{Ma'en Saleh and Liang~Dong,~\IEEEmembership{Senior Member,~IEEE}
\thanks{M.~Saleh is with the Department of Communications and Computer Engineering, Tafila Technical University, Tafila, 66110, Jordan (e-mail: maen@ttu.edu.jo).  }
\thanks{L.~Dong is with the Department of Electrical and Computer Engineering, Baylor University, Waco, TX, 76798 USA (e-mail: liang\_dong@baylor.edu).}
}

\maketitle

\begin{abstract}
Secure routing protocols are proposed for the vehicular ad hoc networks.  The protocols integrate the security authentication process with the Location-Aided Routing (LAR) protocol to support Wi-Fi Direct communications between the vehicles.   The methods are robust against various security threats.   The security authentication process adopts a modified Diffie-Hellman key agreement protocol.   The Diffie-Hellman protocol is used with a short authentication string (SAS)-based key agreement over Wi-Fi Direct out-of-band communication channels.  It protects the communication from any man-in-the-middle security threats.  In particular, the security process is integrated into two LAR routing schemes, i.e., the request-zone LAR scheme and the distance-based LAR scheme.   We conduct extensive simulations with different network parameters such as the vehicular node density, the number of the malicious nodes, and the speed of the nodes.   Simulation results show that the proposed routing protocols provide superior performance in secure data delivery and average total packet delay.  Also, the secure distance-based LAR protocol outperforms the secure request-zone LAR protocol.


\end{abstract}



%
\IEEEpeerreviewmaketitle

\section{Introduction}

Vehicular ad hoc network (VANET) is a class of mobile ad hoc network (MANET), where a group of vehicles with high mobility provide connectivity to each other~\cite{5343061}.   The intercommunication of the vehicular nodes can be through a direct transmission from the source to the destination if the destination is in the sources transmission range.  It can also go through the intermediate nodes if the destination is outside the source's transmission range~\cite{raw2015analytical}.     The intelligent transportation system (ITS) uses the VANET technology to provide safety services to customers such as slow-down notifications, collision warnings, emergency notifications, and road enforcement~\cite{6118326}.   A reliable intercommunication should be established for the safety services, because a disconnection may lead to a catastrophe~\cite{6697835}.  

In a MANET, different routing protocols are adopted to provide reliable communications.  These protocols have methods that are proactive (e.g., OLSR and DSDV), reactive (e.g., AODV and DSR), hybrid (e.g., ZRP), and hierarchical (e.g., CBRR and FSR)~\cite{5426524,6583337,6708418,7546607}.  Although these methods are efficient for MANETs, they do not perform well in VANETs.  This is because that the VANET does not follow a fixed topology or a low-dynamic topology.  When high-mobility vehicular nodes communicate with each other, the relative distances change dramatically, and the connections are intermittent.  Therefore, the network topology is highly dynamic.  It leads to unreliable intercommunication with existing MANET routing protocols.  Also, the traffic restrictions such as intersections, traffic lights, road patterns, and signal blocking objects degrade the reliability of the routes discovered by the MANET routing protocols~\cite{7286854}. 
To overcome the limitations of MANET routing protocols, the topology-independent routing protocols, in particular, the position-based routing protocols such as Location-Aided Routing (LAR), Greedy Perimeter Stateless Routing (GPSR), and Greedy Perimeter Coordinator
Routing (GPCR)~\cite{6060927,7375000,6563165,6684213}, should be adopted for VANETs.   In the proposed routing protocols, the physical locations of the vehicular nodes are found with the GPS service~\cite{6544635}.  The position-based routing protocols provide a stable and reliable route to the destination, where the positions of the nodes are accurate through GPS service.

One of the common communication methods for the VANET is the dedicated short range communication (DSRC) that is based on the IEEE802.11p standard.   It provides a reliable communication link between two vehicles.   However, the DSRC technology has limitations in the cost of the additional dedicated hardware, the low channel bandwidth (10 MHz), and the low data rate (6-27 Mbps)~\cite{7401382}.   The Wi-Fi Direct technology based on the IEEE 802.11n standard is used to establish a device-to-device (D2D) connection without the coordination of the access point~\cite{7579020}.  It has been proven with real-time experiments that the Wi-Fi is a successful means of communication between vehicles in a VANET even at very high speed (i.e., 120 mph)~\cite{4537038}.   Compared with DSRC, Wi-Fi Direct provides a channel bandwidth of 20 MHz, up to 250 Mbps data rates, and no additional hardware cost.   The proposed routing protocols work with Wi-Fi Direct communication links to provide the required QoS requirements of the real-time data flow in the VANET.

The VANET is subject to many security threats that include altering GPS information, position cheating, identifier altering, snooping and spoofing, and  man-in-the-middle attack (MITMA)~\cite{6472013, 6565569}.   Hacking is a serious problem in the VANET  due to information broadcasting, infrastructure-less model, and high-dynamic topology changes~\cite{7127003,7754846}. Accordingly, various secure routing protocols are proposed for the VANET~\cite{6166905,6512236}.  
The geographical secure path routing (GSPR) protocol adds authentication and privacy to the geographic path routing (GPR) protocol through sharing geographic hashes to detect malicious nodes~\cite{4640905}.   To deal with false-position security threats, digital signature and plausibility checks are used in a vehicle-to-vehicle secure position-based routing protocol~\cite{4349671}.   Hash message authentication code (HMAC) is integrated with the optimized link state routing (OLSR) to generate the secure OLSR (SOLSR) routing protocol~\cite{7425245}. The protocol detects the snooping security threats through the use of symmetric and public cryptographic key.   Integrity security service is provided for the VANET through the secure ad hoc on-demand distance vector (SAODV) routing protocol~\cite{5376147}. The hop count process is applied through the use of a hash chain, while the digital signature is used for authentication.  

In this paper, we propose new routing protocols for the VANET that integrate the LAR with the security authentication process over Wi-Fi Direct data links between the vehicles.  The routing protocols include the request-zone LAR (RLAR) scheme and the distance-based LAR (DLAR) scheme.  The security unit adopts the Diffie-Hellman protocol with a short authentication string (SAS)-based key agreement over Wi-Fi Direct out-of-band communication channels.   The proposed protocols protect the VANET from the security threats, especially the MITMA.  

The key features of the proposed secure routing for the VANET are as follows.
\begin{enumerate}
\item The use of Wi-Fi Direct technology for data communication between vehicle nodes, thus providing the VANET with efficient node intercommunications. 
\item The integration of the two LAR routing protocols with a security-authentication unit to provide a secure route in the VANET.
\item The integration of the Diffie-Hellman protocol with an SAS-based key agreement to provide a high level of security and make the VANET robust against threats such as the MITMA. 
\end{enumerate}

The rest of the paper is organized as follows.  The RLAR protocol is proposed in Section~\ref{sec:RLAR}.  We describe the insecure Wi-Fi Direct RLAR and present the secure Wi-Fi Direct RLAR methods.  The DLAR protocol is proposed in Section~\ref{sec:DLAR}.  We describe the insecure Wi-Fi Direct DLAR and present the secure Wi-Fi Direct DLAR methods.   With extensive simulations, Section~\ref{sec:simulation} gives the performance evaluation of the proposed routing protocols.  Finally, conclusions are drawn in Section~\ref{sec:conclusion}.

\section{Request-zone LAR (RLAR) Protocol}
\label{sec:RLAR}

\subsection{Expected Zone and Request Zone}

Suppose that each of the vehicular network nodes has its location information through the GPS service.
The Expected Zone is a region in which the source (S) node expects the destination (D) node to be contained at some particular time~\cite{ko2000location}.   Assume that node S knows that node D is at location L at time $t_{0}$ and it travels at an average speed of $v$.  From the viewpoint of S, the expected zone of node D at time $t_{1}$ is the circular region of radius $v(t_{1} - t_{0})$ centered at point $L$.  


Node S knows its current location $(X_{s},Y_{s})$.  It also knows the location of node D at time $t_{0}$, i.e., $(X_{d}, Y_{d})$, and the average speed $v$ of D.  Such information can be obtained by the auto-reply messages from the nodes.  Node S wants to communicate with node D at time $t_{1}$.  Accordingly, node S perceives the expected zone of node D at time $t_{1}$ as a circular region with radius $R = v(t_{1}-t_{0})$ and centered at location $(X_{d}, Y_{d})$.

\begin{figure}[!t]
\centering
\includegraphics[width=8.2cm]{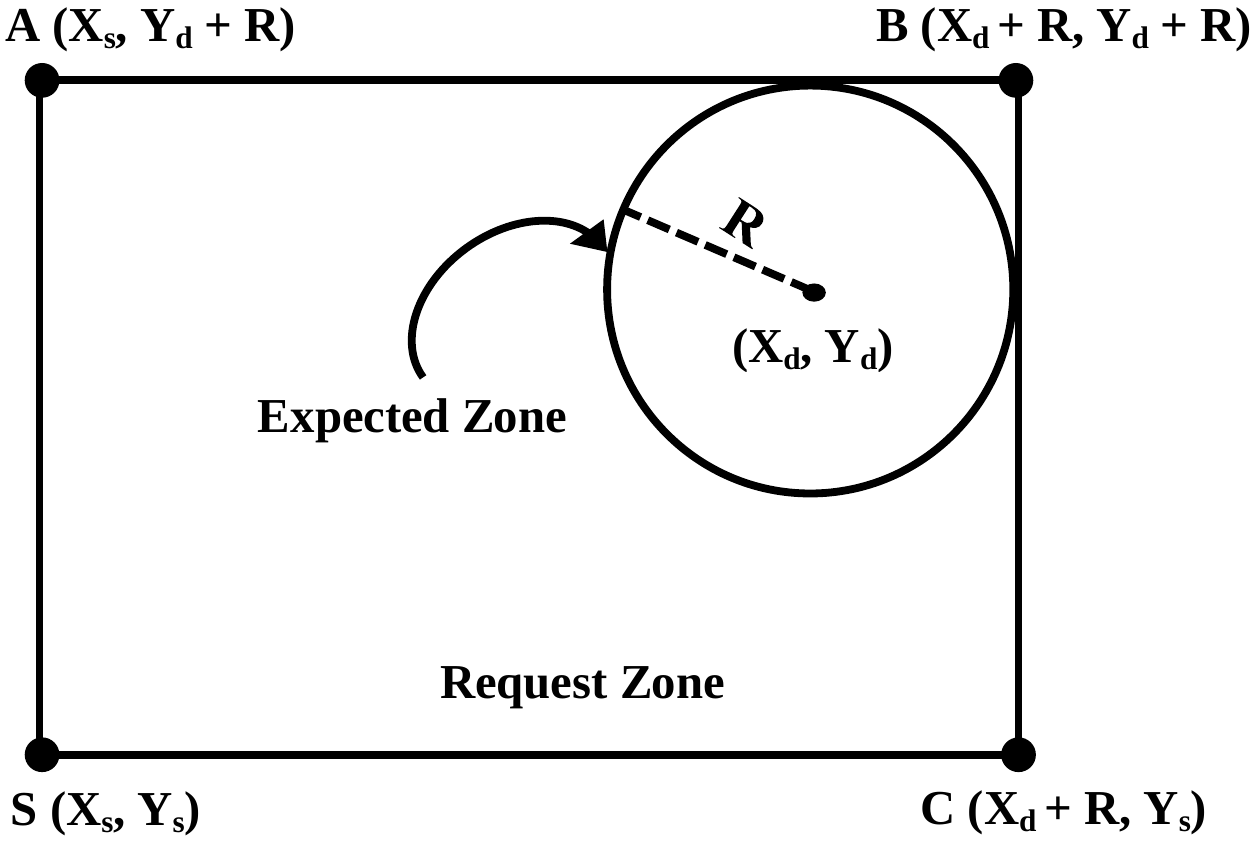}
\caption{Request zone when $Dist > R$.  Source node S is out of the expected zone of destination node D.}
\label{fig:requestzone1}
\end{figure}

\begin{figure}[!t]
\centering
\includegraphics[width=7.6cm]{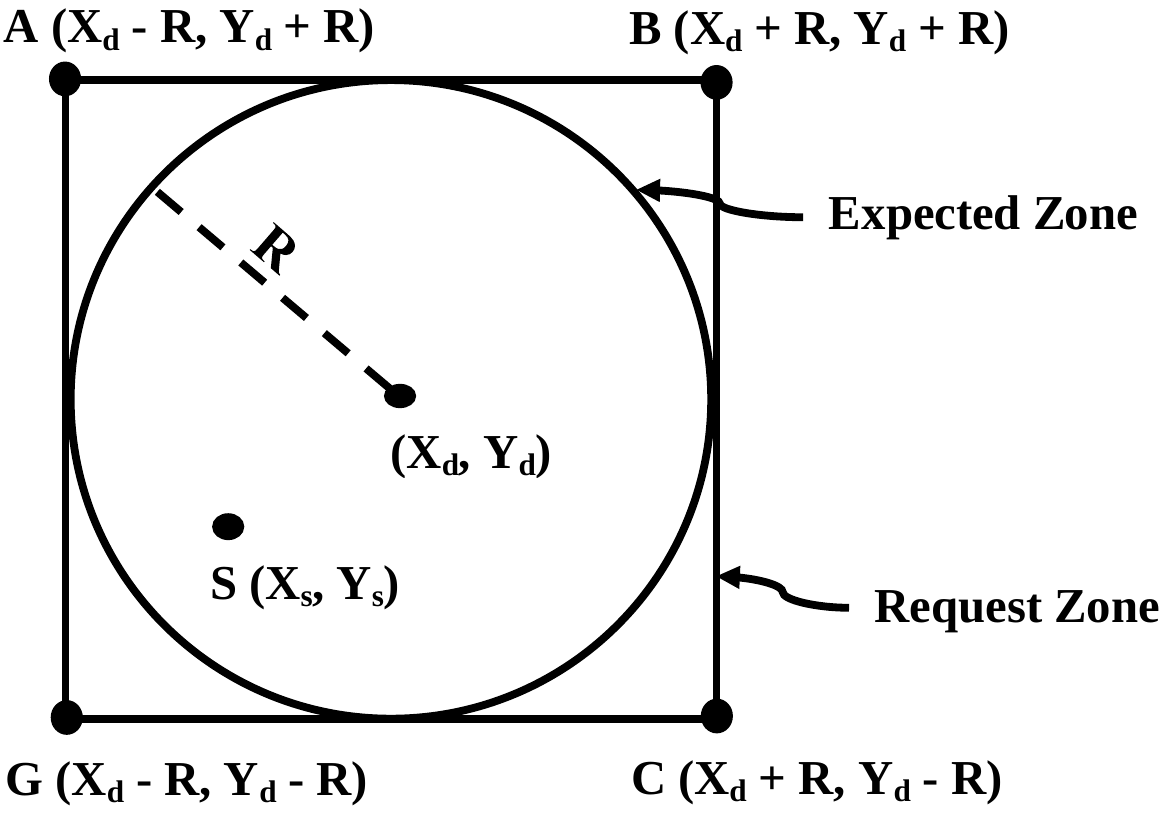}
\caption{Request zone when $Dist \leq R$.  Source node S is in the expected zone of destination node D.}
\label{fig:requestzone2}
\end{figure}

Node S evaluates the distance ($Dist$) between its location $(X_{s}, Y_{s})$ and node D's location $(X_{d}, Y_{d})$.   With  $Dist$, node S defines the Request Zone for the route request.  It is the smallest rectangle that includes the current location of S and the expected zone of D such that the sides of the rectangle are parallel to the X and Y axes.  Given $Dist$ and $R$, there are two situations:\\
(1) If $Dist > R$, node S is out of the expected zone of node D.  The request zone coordinates are shown in Fig.~\ref{fig:requestzone1}.\\
(2) If $Dist \leq R$, node S is in the expected zone of node D.  The request zone coordinates are shown in Fig.~\ref{fig:requestzone2}.

\subsection{RLAR with Wi-Fi Direct Communications}

The LAR protocol depends on the nodes' location information to calculate the route from the source to the destination.   We propose the request-zone LAR (RLAR) approach that uses the request zone information in routing.

In the RLAR protocol, Node S prepares the message for node D that includes: (1) Source coordinates $(X_{s}, Y_{s})$, (2) The coordinates of the request zone (S, A, B, and C in Fig.~\ref{fig:requestzone1}) or (G, A, B, and C in Fig.~\ref{fig:requestzone2}),  (3) Source and destination MAC addresses (e.g., vehicle plate numbers or engine serial numbers in a VANET), and (4) The position of S corresponding to the request zone (e.g. inside or outside the request zone). Node S floods the message to its neighbors with the Wi-Fi Direct links.  When a neighboring node B receives the message, it checks whether its location $(X_{b}, Y_{b})$ is within the request zone. If node B is within the request zone, we have
\begin{eqnarray*}
X_{s} \leq X_{b} \leq (X_{d} + R) & \& & Y_{s} \leq Y_{b} \leq (Y_{d} +R) \\
(X_{d} - R) \leq X_{b} \leq (X_{d} + R) & \& & Y_{d}-R \leq Y_{b} \leq Y_{d}+R
\end{eqnarray*}

If node B is within the request zone, it checks whether the destination address is its address.  If not, node B forwards the message to its neighbors.  If node B is the destination, it generates an ACK message, i.e., a reply, and floods it.  The reply message contains information about the current time and the destination node's speed.  Such information will be used by the source node for defining the request zone for future communication.  If node B does not belong to the request zone, it discards the message.  Also, if a node receives the same message from a different node, it discards it.  This protects the network from being congested.  

\begin{figure}[!t]
\centering
\includegraphics[width=8.8cm]{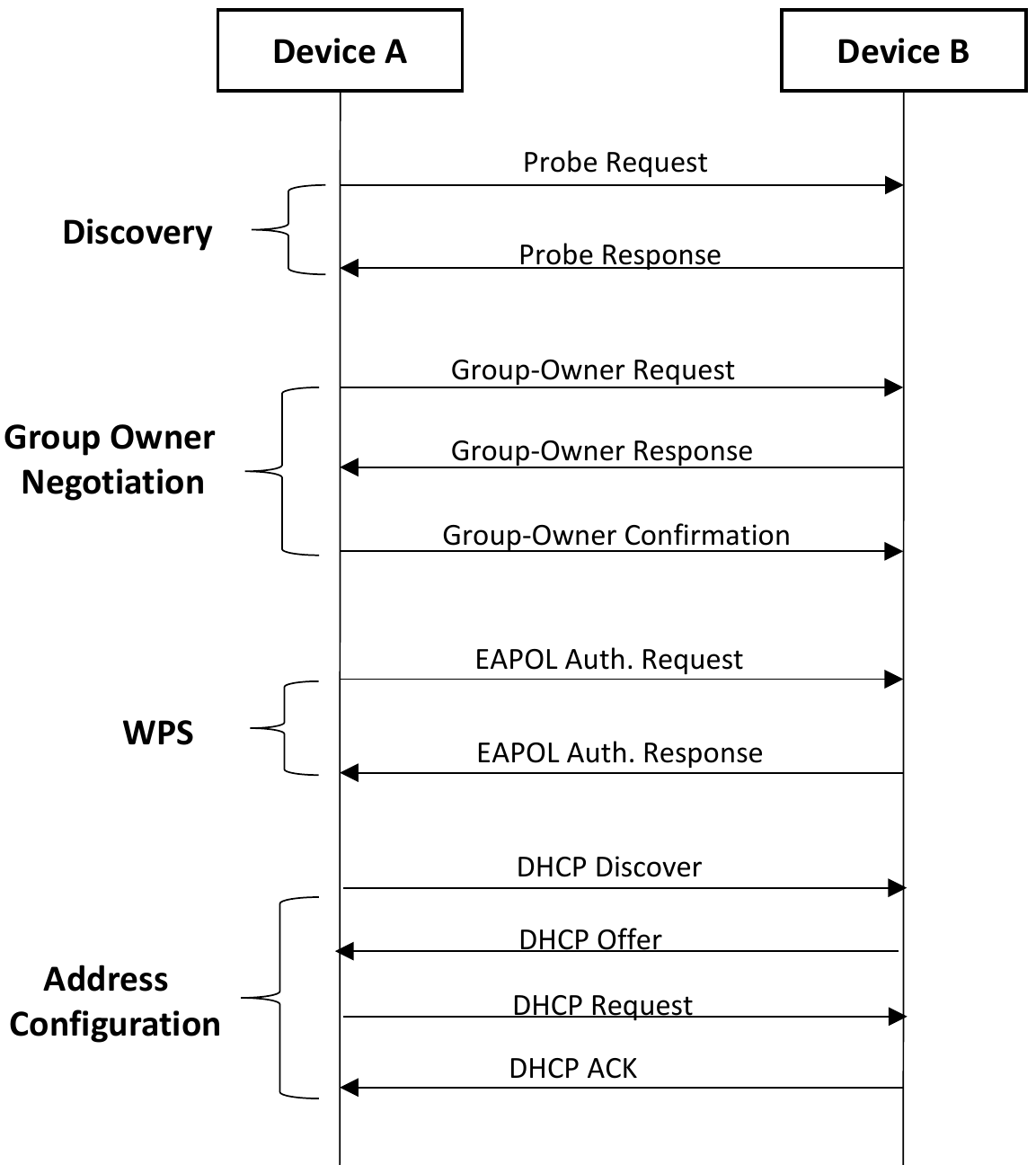}
\caption{Communication protocol with Wi-Fi Direct.}
\label{fig:protocol_WiFiDirect}
\end{figure}

Between any two vehicular nodes of the VANET, the Wi-Fi Direct technology is used as a communication means.   Fig.~\ref{fig:protocol_WiFiDirect} shows the communication protocol between a pair of nodes with Wi-Fi Direct.  The first phase is the discovery process, where the two nodes perform channel probing mechanism with the probe request and probe response control signals.  In the second phase, the group owner is negotiated through group-owner request, response, and confirmation.   Once the group owner is specified, it acts as an access point for the connection.  In the third phase, the Wi-Fi protected setup (WPS) is initiated by the group owner using the extensible authentication protocol signals such as EAPOL request and EAPOL response.  Finally, the address configuration phase is initiated by the group owner by conducting the Dynamic Host Configuration Protocol (DHCP)~\cite{7579020}.

\subsection{Secure RLAR with Wi-Fi Direct Communications}

According to the proposed protocol, we assume the following: 
\begin{enumerate}
\item Any two trusted adjacent nodes belonging to the VANET can initiate an out-of-band channel.  This channel is a trusted one such that it cannot be manipulated by the attackers; 
\item For generating a shared security key, the trusted nodes that belong to the VANET use the Diffie-Hellman protocol with the common integer parameters such as  the prime modulus ($m$) and the base ($b$);
\item Each node $N_{i}$ has its own private key ($r_{i}$), which is an integer not be exchanged. This key will be used to generate the public key in the network; 
\item Each node $N_{i}$ has its unique identification (ID) that can be considered as the MAC address in the network (e.g., the vehicle plate number);
\item Each node $N_{i}$ can generate a $k$-bit random string ($A_{i}$).   These $k$ bits will be used to generate the authentication string ($S_{i}$) of the short authentication string (SAS)-based key agreement protocol.
\end{enumerate}

The Diffie-Hellman key agreement allows two vehicular nodes with no prior knowledge of each other to jointly establish a shared secret key~\cite{7579020}.  The short authentication string (SAS)-based key agreement protocol involves minimal mutual authentication.  It utilizes a cryptography commitment scheme.  An efficient construction of a commitment is achieved by using a cryptographic hash function.   Both nodes compute a hash value of the obtained shared key and compare the hash values via the secure out-of-band channel.  

The source node S performs the Diffie-Hellman key agreement protocol to generate its public key $g_{s}$ as 
\begin{equation}\label{eq:publickey}
g_{s} = b^{r_{s}}  \text{~mod~}  m.
\end{equation}

The source node S then generates a message $m_{s}$ in the form of the concatenation of its public key $g_{s}$ and the randomly generated $k$-bit string $A_{s}$ as
\begin{equation}\label{eq:concatenation}
m_{s} = g_{s} \| A_{s}.
\end{equation}

In our proposed protocol, we use the commitment scheme of the cryptography schemes.   On the one hand, a node is committed to a value and keeps it hidden from others (commit phase).   On the other hand, it has the ability to unlock and reveal such a value later (open phase).   An efficient construction of the commitment scheme can be achieved by using a cryptographic hash function~\cite{pass2003deniability}.  

The source node S uses its private key $r_{s}$ with a cryptographic hash function $H$ to compute the commitment $c_{s}$ on the concatenation $m_{s}$ as 
\begin{equation}\label{eq:commitment}
c_{s} = H (m_{s},r_{s}).
\end{equation}
The source node S then includes the following information in the message that is sent to the destination: (1) The commitment $c_{s}$, (2) The source coordinates ($X_{s}$, $Y_{s}$), (3) The request zone coordinates (S A B C in Fig.~\ref{fig:requestzone1}) or (G A B C in Fig.~\ref{fig:requestzone2}), and (4) Source and destination MAC addresses (ID$_{s}$, ID$_{d}$).  The source node S floods the message to its neighbors with Wi-Fi Direct. 

When a node N receives the message, it checks whether its location ($X_{n}$, $Y_{n}$) is within the request zone.  If the node N is not within the request zone, it discards the message.  Otherwise, it generates its own concatenation $m_{n}$ using formulas that are similar to \eqref{eq:publickey} and \eqref{eq:concatenation}.
Node N then sends $m_{n}$ to the source node S (with the destination address ID$_{s}$).   Once the source node S receives this message, it sends the open parameter $w$ to node N (with destination address ID$_{n}$).   

Before generating the shared security key, node S generates the $k$-bit authentication string $S_{s}$ as
\begin{equation}\label{eq:auth_string_Ss}
S_{s} = A_{s} \oplus A_{n} 
\end{equation}
where $A_{n}$ is extracted from $m_{n}$.  
Node N uses the open parameter $w$ to reveal the commitment $c_{s}$ and extracts the $k$-bit string $A_{s}$ from $m_{s}$. 
Node N then generates the $k$-bit authentication string $S_{n}$ as
\begin{equation}\label{eq:auth_string_Sn}
S_{n} = A_{s} \oplus A_{n}.
\end{equation}

\begin{figure}[!t]
\centering
\includegraphics[width=8.8cm]{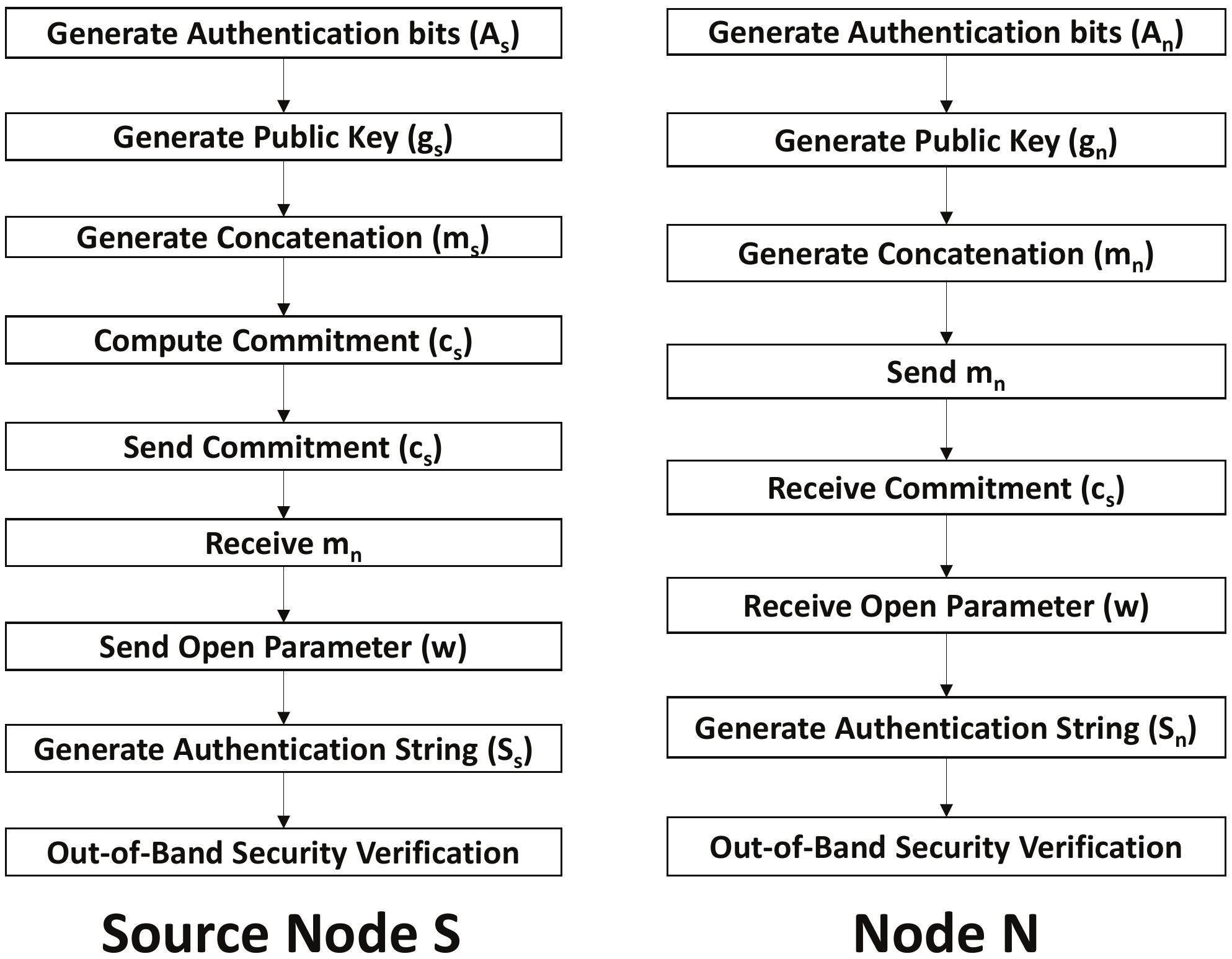}
\caption{Security authentication model at source node S and node N.}
\label{fig:security_authentication}
\end{figure}

Over the secure out-of-band channel, nodes S and N verify whether the two authentication strings match ($S_{s} = S_{n}$?). The overall security-authentication model at the two parties (source node S and node N) is shown in Fig.~\ref{fig:security_authentication}.

If the two authentication strings do not match, the two parties stop the process of generating the security keys, and node N discards the message due to an MITMA. Therefore, node N will not be in the secure route to the destination.  The source node S will use another adjacent node for the secure route.  If the two strings match, both nodes S and N generate the shared key as 
\begin{eqnarray}
\mathrm{Key}(s) &=& (g_{n})^{r_{s}}\mathrm{~mod~} m \label{eq:key_s}\\
\mathrm{Key}(n) &=& (g_{s})^{r_{n}} \mathrm{~mod~} m \label{eq:key_n}
\end{eqnarray}
where $\mathrm{Key}(s)$ and $\mathrm{Key}(n)$ are the shared security keys at node S and node N, respectively. These two values are equal, i.e., $\mathrm{Key}(s) = \mathrm{Key}(n)$. Note that, nodes S and N do not share such keys.  They generate them using the shared public keys $g_{s}$  and $g_{n}$. 

\begin{figure}[!t]
\centering
\includegraphics[width=7cm]{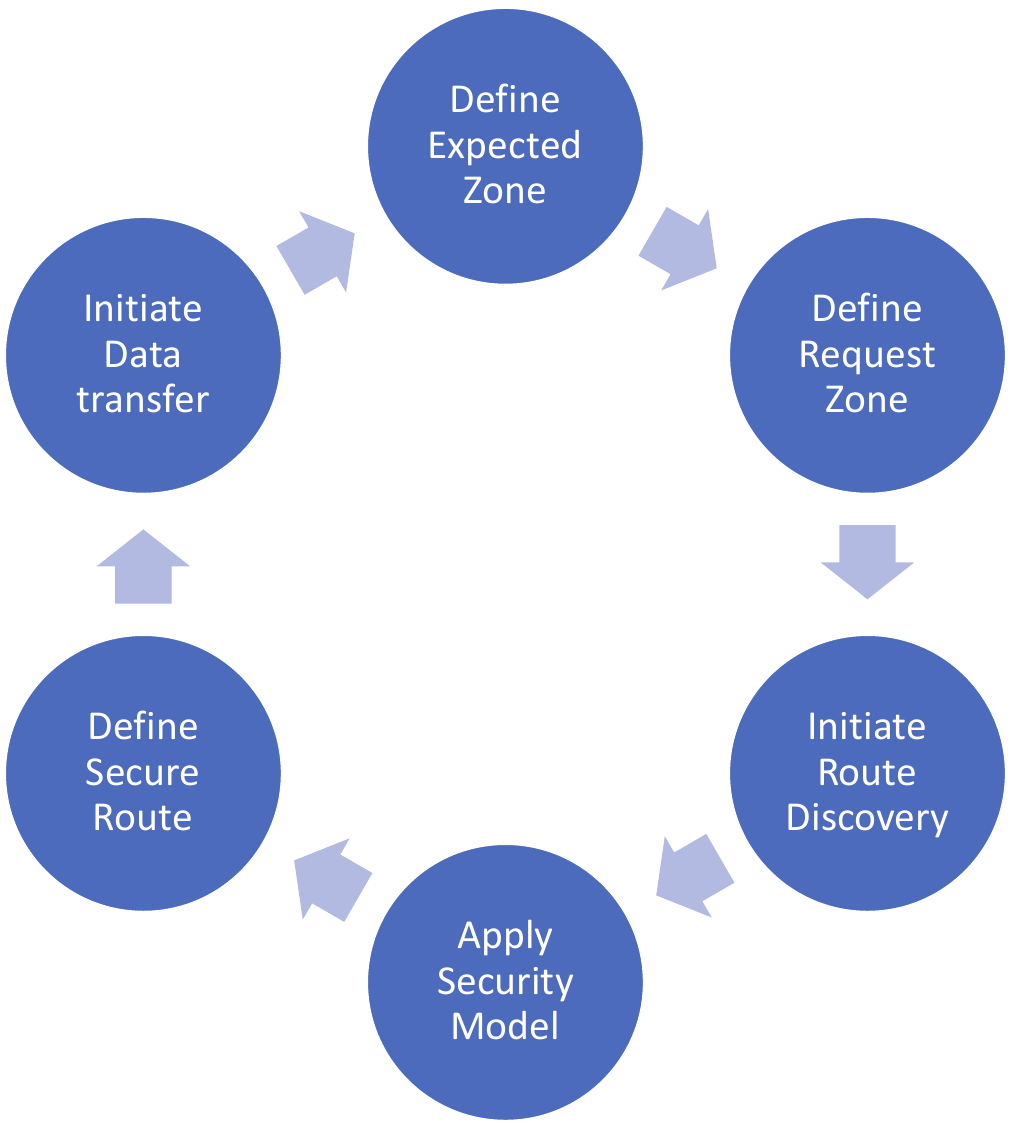}
\caption{Six phases of the secure RLAR protocol.}
\label{fig:phases_RLAR}
\end{figure}

Node N checks whether the destination address is its address.  If not, it forwards the message that includes the request area coordinates to its neighbors.  The same security authentication procedure repeats.  If yes, node N generates an ACK message (reply) and floods it.  The reply message contains information about the current time and the destination's speed.  Such information will be used by the source for defining a new request zone for future communication.  If a node receives the same message from a different node, it discards it.  This process protects the network from being congested. The phases of the proposed secure RLAR routing protocol is shown in Fig.~\ref{fig:phases_RLAR}.

\section{Distance-based LAR (DLAR) Protocol}
\label{sec:DLAR}

\subsection{DLAR with Wi-Fi Direct Communications}

\begin{figure}[!t]
\centering
\includegraphics[width=5cm]{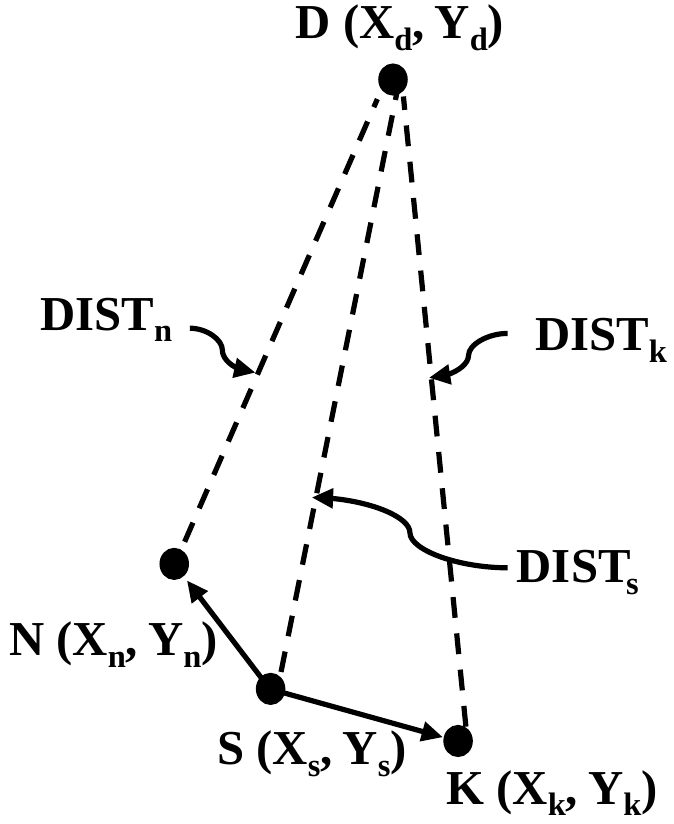}
\caption{Distance-based LAR (DLAR) scheme.}
\label{fig:DLAR}
\end{figure}

According to the above RLAR protocol, the route calculation is restricted by the boundaries of the requested zone.  This may cause successive route disconnection.  To overcome this limitation, we propose the distance-based LAR (DLAR)~\cite{ko2000location}.  As shown in Fig.~\ref{fig:DLAR}, the only restriction of the route calculation is the node's transmission range.

Assume the source node S knows the following information:
\begin{enumerate}
\item Its current location ($X_{s}$, $Y_{s}$) via GPS; 
\item The location ($X_{d}$, $Y_{d}$) of the destination node D at some time $t_{0}$.
\end{enumerate}
The route discovery is initiated by node S at time $t_{1}$, where $t_{1} \geq t_{0}$.  The source node S calculates its distance from the location ($X_{d}$, $Y_{d}$), which is denoted as $\mathrm{DIST}_{s}$.  It includes in the message sent to node D (1) $\mathrm{DIST}_{s}$, (2) the coordinates of node D ($X_{d}$, $Y_{d}$), and (3) source and destination MAC addresses, e.g., vehicle plate numbers.  The source node S floods the message to its neighbors through Wi-Fi direct.  

When a node N receives the message, it calculates its distance $\mathrm{DIST}_{n}$ from the destination coordinates ($X_{d}$, $Y_{d}$) and compares $\mathrm{DIST}_{n}$ with $\mathrm{DIST}_{s}$.
If $\mathrm{DIST}_{n} \leq \mathrm{DIST}_{s}$, node N belongs to the route to the destination.  Node N checks whether itself is the destination node.  If it is not, node N replaces $\mathrm{DIST}_{s}$ with $\mathrm{DIST}_{n}$ in the message and forwards the message to its neighbors.   If node N is the destination node, it generates an ACK message (reply) and floods it.  This operation repeats until the message is received by the destination node.  

If a node receives the same message from a different node, it discards it.  This protects the network from being congested.

\subsection{Secure DLAR with Wi-Fi Direct Communications}

We assume the following when integrating the security module in the DLAR.
\begin{enumerate}
\item The source node S knows its current location ($X_{s}$, $Y_{s}$) via GPS;
\item The source node S has knowledge about the location of node D ($X_{d}$, $Y_{d}$) at some time $t_{0}$.  Route discovery is initiated by node S at time $t_{1}$ with ($t_{1} \geq t_{0}$); 
\item Each two trusted adjacent nodes in the VANET can establish an out-of-band channel.   This channel is trusted and cannot be manipulated by the attackers;
\item The vehicular nodes use the Diffie-Hellman protocol to generate a shared security key with the common integer parameters such as the prime modulus (m) and the base (b); 
\item Each node $N_{i}$ has its own private key ($r_{i}$), which is an integer not to be exchanged.  This key will be used to generate the public keys in the network; 
\item Each node has its unique identification (ID) that can be considered as the MAC address in the network (e.g., the vehicle plate number); 
\item Each node ($N_{i}$) can generate $k$-bit random string ($A_{i}$). This $k$-bit string will be used to generate the authentication string ($S_{i}$) of the short authentication string (SAS)-based key agreement protocol.
\end{enumerate}

Accordingly, the source node S performs the Diffie-Hellman key agreement protocol to generate its public key $g_{s}$ as in \eqref{eq:publickey}.  The source node S then makes the concatenation $m_{s}$ of its public key $g_{s}$ and the randomly generated $k$-bit string $A_{s}$ as in \eqref{eq:concatenation}.  The source node S uses its private key $r_{s}$ with a cryptographic hash function $H$ to compute the commitment $c_{s}$ on the concatenation $m_{s}$ as in \eqref{eq:commitment}.

For the secure DLAR, the source node S calculates its distance from location ($X_{d}$, $Y_{d}$) which is denoted by $\mathrm{DIST}_{s}$.  It includes the following information in the message that is sent to the destination: (1) The commitment $c_{s}$, (2) $\mathrm{DIST}_{s}$, (3) the coordinates of the destination node D ($X_{d}$, $Y_{d}$), and (4) source and destination MAC addresses ($\mathrm{ID}_{s}$, $\mathrm{ID}_{d}$).  The source node floods the message to its neighbors using Wi-Fi Direct.  

When a node N receives the message, it calculates its distance ($\mathrm{DIST}_{n}$) from the destination ($X_{d}$, $Y_{d}$).  Node N determines whether itself should be in the route by comparing $\mathrm{DIST}_{n}$ with $\mathrm{DIST}_{s}$.  If $\mathrm{DIST}_{n} > \mathrm{DIST}_{s}$, it reasons that it is not in the route and discards the received message.  If $\mathrm{DIST}_{n} \leq \mathrm{DIST}_{s}$, it knows it belongs to the route and generates its concatenation $m_{n}$.  

Node N sends $m_{n}$ to the source node S with the destination address $\mathrm{ID}_{s}$.   Once the source S receives such message, it sends the open parameter $w$ to node N  with the destination address $\mathrm{ID}_{n}$. 

Before generating the shared security key, node S and node 
N generate the authentication string using \eqref{eq:auth_string_Ss} and \eqref{eq:auth_string_Sn}, respectively.   Note that, node N uses the open parameter $w$ to reveal the commitment $c_{s}$ and extracts the $k$-bit string $A_{s}$ from $m_{s}$.

\begin{figure}[!t]
\centering
\includegraphics[width=8.8cm]{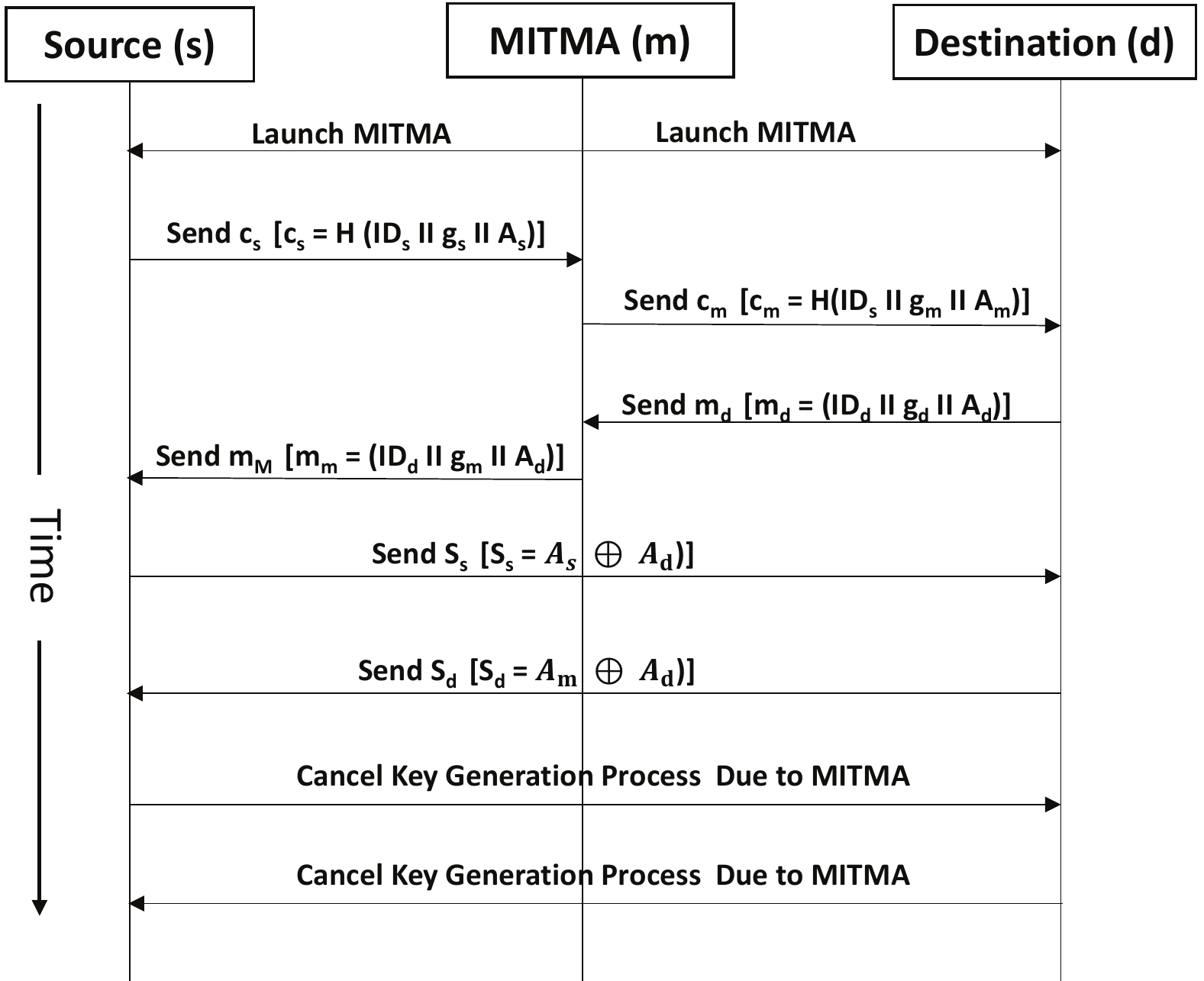}
\caption{Discovery of the man-in-the-middle attack (MITMA) security threat.}
\label{fig:discover_MITMA}
\end{figure}

Over the secure out-of-band channel, node S and node N verify whether the two authentication strings match ($S_{s} = S_{n}$?).  If the strings do not match, the two parties stop the process of generating the security keys.  Node N discards the message due to an MITMA.   Node N is not in the secure route to the destination, and node S will use another adjacent node for a secure route.  If the two strings match, both node S and node N generate the shared keys $\mathrm{Key}(s)$ and $\mathrm{Key} (n)$ according to \eqref{eq:key_s} and \eqref{eq:key_n}, respectively.  These two values are equal.  Note that, the nodes do not share such keys. They generate them using the shared public keys $g_{s}$ and $g_{n}$.  

Node N checks whether the destination address is its address. If not, it forwards the message to its neighbors.  The message includes the destination's coordinates ($X_{d}$, $Y_{d}$) and the distance $\mathrm{DIST}_{n}$ instead of $\mathrm{DIST}_{s}$ .  The operation repeats until reaching the destination.   If node N is the destination node, it generates an ACK message (reply) and floods it. 

If a node receives the same message from a different node, it discards the message.  Such process protects the network from being congested. 

\begin{figure}[!t]
\centering
\includegraphics[width=7cm]{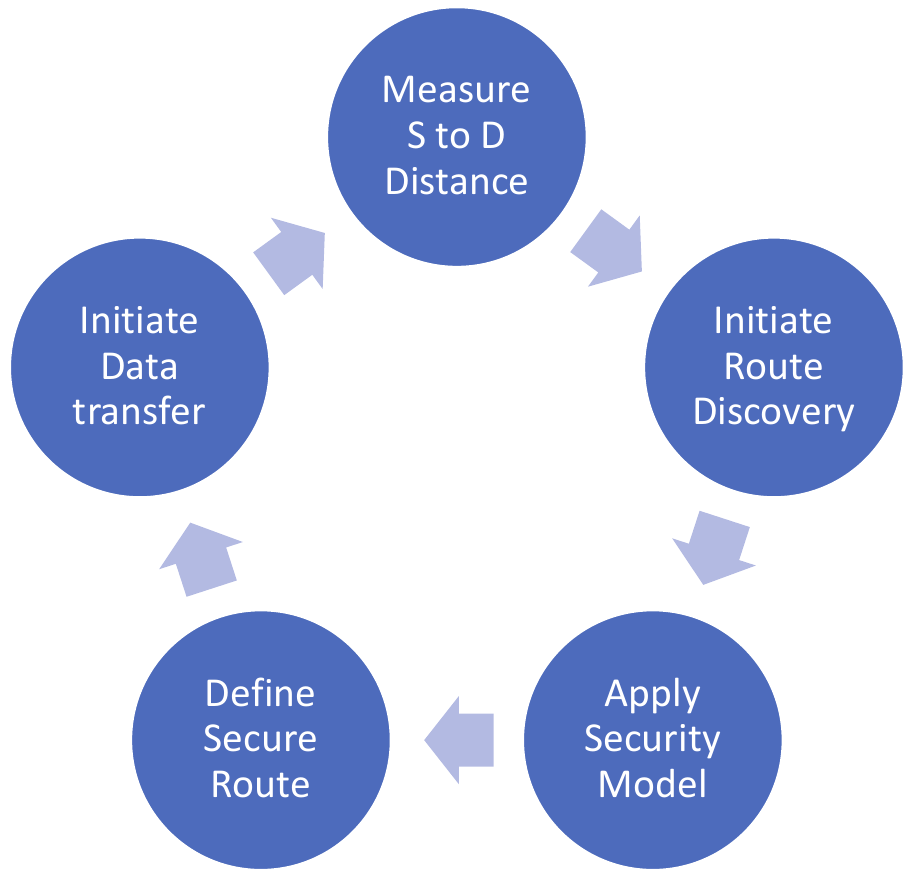}
\caption{Five phases of the secure DLAR protocol.}
\label{fig:DLAR_phases}
\end{figure}

The overall process of discovering the MITMA is shown in Fig.~\ref{fig:discover_MITMA}.
The five phases of the proposed secure DLAR is shown in Fig.~\ref{fig:DLAR_phases}.

\section{Simulation Results}
\label{sec:simulation}

Because the VANET environment is heterogeneous and dynamic, conventional network simulators are not sufficient for analyzing the real-time performance of the proposed routing protocols with security aspects.  Besides, conventional network simulators such as ns-3 do not include geographic routing in their standard codes.   They work well for wireless networks and MANET, but not for the VANET under consideration.  In this research, we build our simulation models based on the .Net platform with its inherited object-oriented capabilities to analyze the performance of the proposed routing protocols.

\subsection{Motion Model, System Parameters and Assumptions}
In the simulations, we set the motion model, system parameters and assumptions as follows.
\begin{enumerate}
\item The number of vehicular nodes is set to be $10, 20, 30, \ldots, N$.  In each simulation run, there are $N/2$ pairs of peer-to-peer communications, where the $i$th node is the source node and the $(i+N/2)$th node is the destination node, $i = 1, 2, \ldots, N/2$.
\item The sending rate $\lambda$ is 50 packets per second, with each node sending for 10 seconds (a total of 500 packets) to the destination.
The inter-arrival time is exponentially distributed with a mean of $1/\lambda$.
\item The initial locations of the nodes $\{(X, Y)\}$ are randomly chosen.
The nodes move continuously with velocity $v$ that is uniformly distributed in $[2, 40]$ units/sec.
The nodes move in a square region of [1000 units $\times$ 1000 units], where the direction of each node's movement is random.
\item Each node randomly changes its direction after traveling a distance of $d$ that is exponentially distributed with a mean of 25 units. 
When a node touches the region boundaries, it will ``bounce back'' and travel the remaining distance in the opposite direction.
\item Transmission range for each node is set to be 200 units.  
\item For each simulation, the Diffie-Hellman security integer parameters, i.e., modulus $m$ and base $b$, are randomly chosen and common to all the nodes.
The private key $r$ is randomly chosen for each individual node,
 and the $k$-bit string $A$ is randomly generated by each node, where $k=10$.
\end{enumerate}

\subsection{Node Density Effect}

We study the performance of the proposed secure routing protocols regarding secure data delivery and average total packet delay on the node density.  The number of nodes in the VANET region is chosen as $N = 10, 20, 30, \ldots, 100$.   The nodes' speed is set to be 5 units/sec.   The percentage of malicious nodes that cause an MITMA is 10\%.  

\begin{figure}[!t]
\centering
\includegraphics[width=0.98\linewidth]{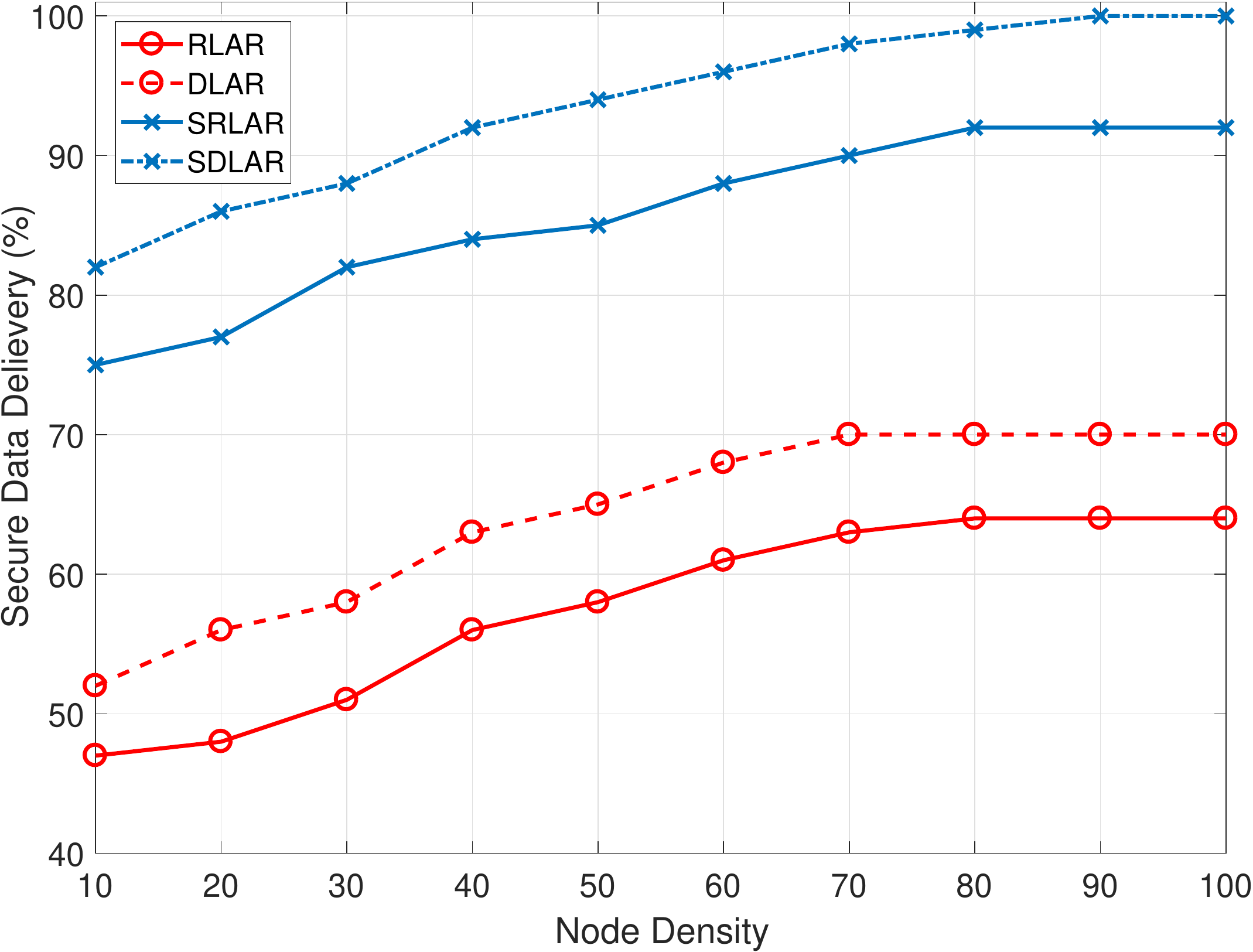}
\caption{Effect of node density on secure data delivery. 10\% malicious nodes. Node speed is 5 units/sec.}
\label{fig:density_delivery}
\end{figure}

Fig.~\ref{fig:density_delivery} shows the percentage of secure data delivery versus the number of nodes.  Integrating the security module to the standard LAR protocol enhances the delivery of the data packets at the destination nodes.  This is because the chances of dropping a packet due to the MITMA are reduced.  Of the two secure routing protocols, the simulation results show that the secure DLAR protocol outperforms the secure RLAR protocol regarding secure data delivery.  This is because the secure DLAR is only limited by the transmission ranges of the nodes whereas the secure RLAR is restricted by both the transmission ranges and the defined expected zones.   Accordingly, the secure DLAR experiences fewer disconnections of the routes hence higher data delivery percentage.

\begin{figure}[!t]
\centering
\includegraphics[width=0.98\linewidth]{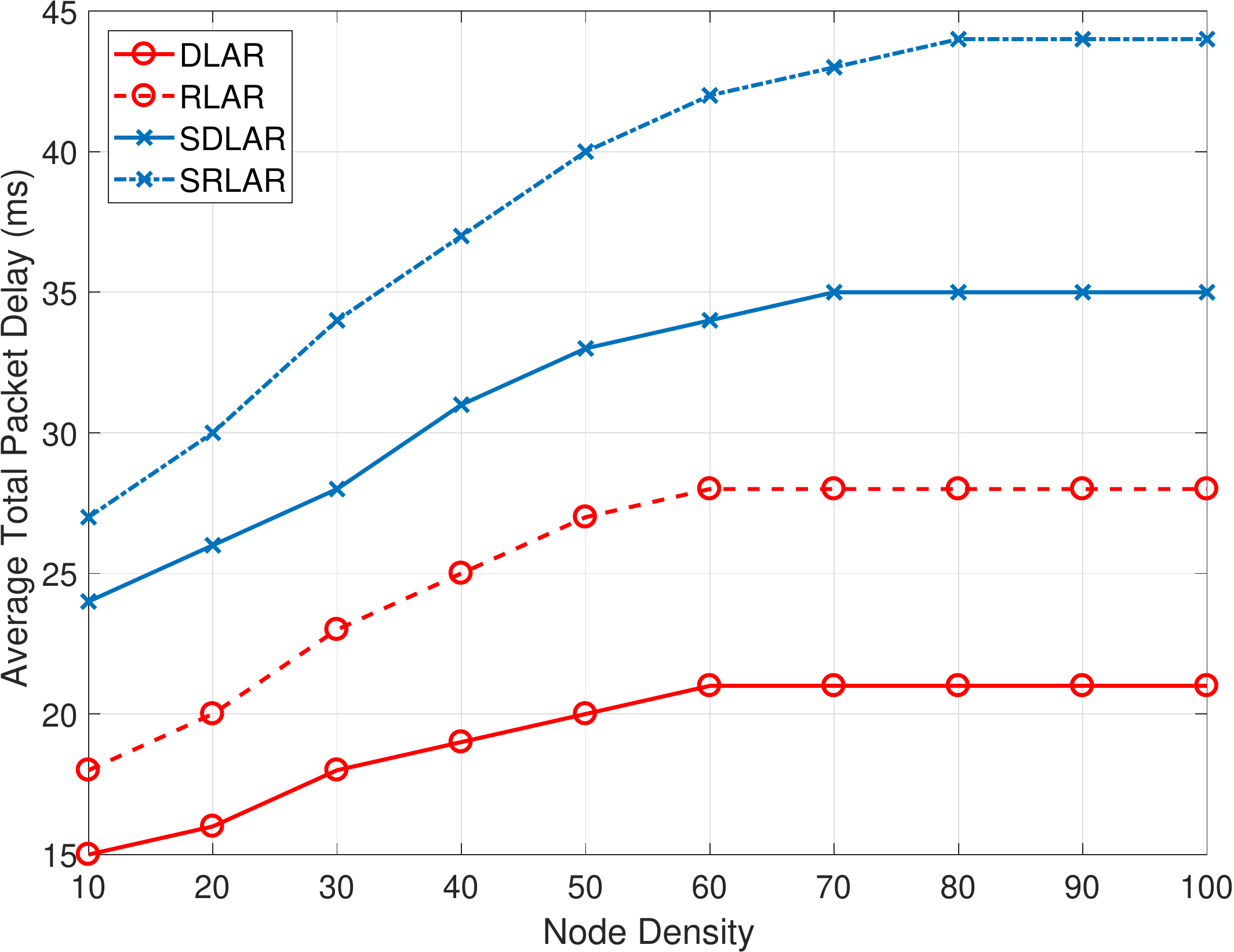}
\caption{Effect of node density on average total packet delay. 10\% malicious nodes.  Node speed is 5 units/sec.}
\label{fig:density_delay}
\end{figure}

Fig.~\ref{fig:density_delay} shows the average total packet delay versus the number of nodes.  The average total packet delay reflects the efficiency of finding a route from the source to the destination with the routing protocol.  The simulation results show that the proposed secure routing protocols have a tradeoff of larger delays compared with the non-secure protocols.  This is because the secure routing protocols have additional security association phases.  Of the two proposed secure routing protocols, secure DLAR outperforms the secure RLAR regarding average total packet delay.  Without the expected zone restriction, the secure DLAR experiences fewer route disconnections.  

To clarify the results, the data are analyzed in the following tables.  Table~\ref{table:delivery_DLAR} and Table~\ref{table:delivery_RLAR} show the enhancement of the data delivery when integrating the security module in the LAR protocol.   It is revealed that the secure DLAR improves the data delivery by an average of 46.2\% over the non-secure DLAR and the secure RLAR improves by an average of 49.6\% over the non-secure RLAR.
Table~\ref{table:delivery_SRLARSDLAR} and Table~\ref{table:delay_SRLARSDLAR} compare the secure DLAR with the secure RLAR.  It is revealed that the secure DLAR enhances the data delivery by an average of 9.14\% and reduces the average total packet delay by an average of 17.48\% over the secure RLAR.

\begin{table}[!b]
\caption{Delivery (DLAR vs. SDLAR)} 
\centering 
\begin{tabular}{c c c c} 
\hline\hline 
\# of Nodes & Delivery (DLAR) & Delivery (SDLAR) & Enhancement\\ [0.5ex] 
\hline 
10&	52&	82&	57.7\%\\%
 20&56&	86&	53.6\%\\%
 30&58&	88&	51.7\%\\%
40	&63&	92&	46\%\\%
50	&65&	94&	44.6\%\\%
60	&68&	96&	41.2\%\\%
70	&70&	98&	40\%\\%
80	&70&	99&	41.4\%\\%
90	&70&	100&	42.9\%\\%
100	&70&	100&	42.9\%\\%
\hline
 &  &  & Avg.=\text{~}46.2\% \\ [1ex] 
\end{tabular}
\label{table:delivery_DLAR} 
\end{table}

\begin{table}[!b]
\caption{Delivery (RLAR vs. SRLAR)} 
\centering 
\begin{tabular}{c c c c} 
\hline\hline 
\# of Nodes & Delivery (RLAR) & Delivery (SRLAR) & Enhancement\\ [0.5ex] 
\hline 
10&	47&	75&	59.6\%\\
20&	48&	77&	60.4\%\\
30&	51&	82&	60.8\%\\
40&	56&	84&	50\%\\
50&	58&	85&	46.6\%\\
60&	61&	88&	44.3\%\\
70&	63&	90&	42.9\%\\
80&	64&	92&	43.8\%\\
90&	64&	92&	43.8\%\\
100&64&	92&	43.8\%\\
\hline
 &  &  & Avg.=\text{~}49.6\% \\ [1ex] 
\end{tabular}
\label{table:delivery_RLAR} 
\end{table}

\begin{table}[!b]
\caption{Delivery (SRLAR vs. SDLAR)} 
\centering 
\begin{tabular}{c c c c} 
\hline\hline 
\# of Nodes & Delivery (SRLAR) & Delivery (SDLAR) & Enhancement\\ [0.5ex] 
\hline 
10&	75&	82&	9.3\%\\
20&	77&	86&	11.7\%\\
30&	82&	88&	7.3\%\\
40&	84&	92&	9.5\%\\
50&	85&	94&	10.6\%\\
60&	88&	96&	9.1\%\\
70&	90&	98&	8.9\%\\
80&	92&	99&	7.6\%\\
90&	92&	100&	8.7\%\\
100&	92&	100&	8.7\%\\
\hline
 &  &  & Avg.=\text{~}9.14\% \\ [1ex] 
\end{tabular}
\label{table:delivery_SRLARSDLAR} 
\end{table}

\begin{table}[!b]
\caption{Delay (SRLAR vs. SDLAR))} 
\centering 
\begin{tabular}{c c c c} 
\hline\hline 
\# of Nodes & Delay (SRLAR) & Delay (SDLAR) & Enhancement\\ 
& [ms] & [ms] & \\ [0.5ex] 
\hline 
10&	27&	24&	11.1\%\\
20&	30&	26&	13.3\%\\
30&	34&	28&	17.6\%\\
40&	37&	31&	16.2\%\\
50&	40&	33&	17.5\%\\
60&	42&	34&	19\%\\
70&	43&	35&	18.6\%\\
80&	44&	35&	20.5\%\\
90&	44&	35&	20.5\%\\
100&	44&	35&	20.5\%\\
\hline
 &  &  & Avg.=\text{~}17.48\% \\ [1ex] 
\end{tabular}
\label{table:delay_SRLARSDLAR} 
\end{table}

\subsection{Security Threat Effect}

We study the effect of the number of malicious nodes, particularly those cause the MITMA, on both data delivery and packet delay.   The number of malicious nodes is set to be $2,4,6,\ldots,12$.  The number of vehicular nodes is fixed at 40, and the nodes' speed is 5 units/sec.

\begin{figure}[!t]
\centering
\includegraphics[width=0.98\linewidth]{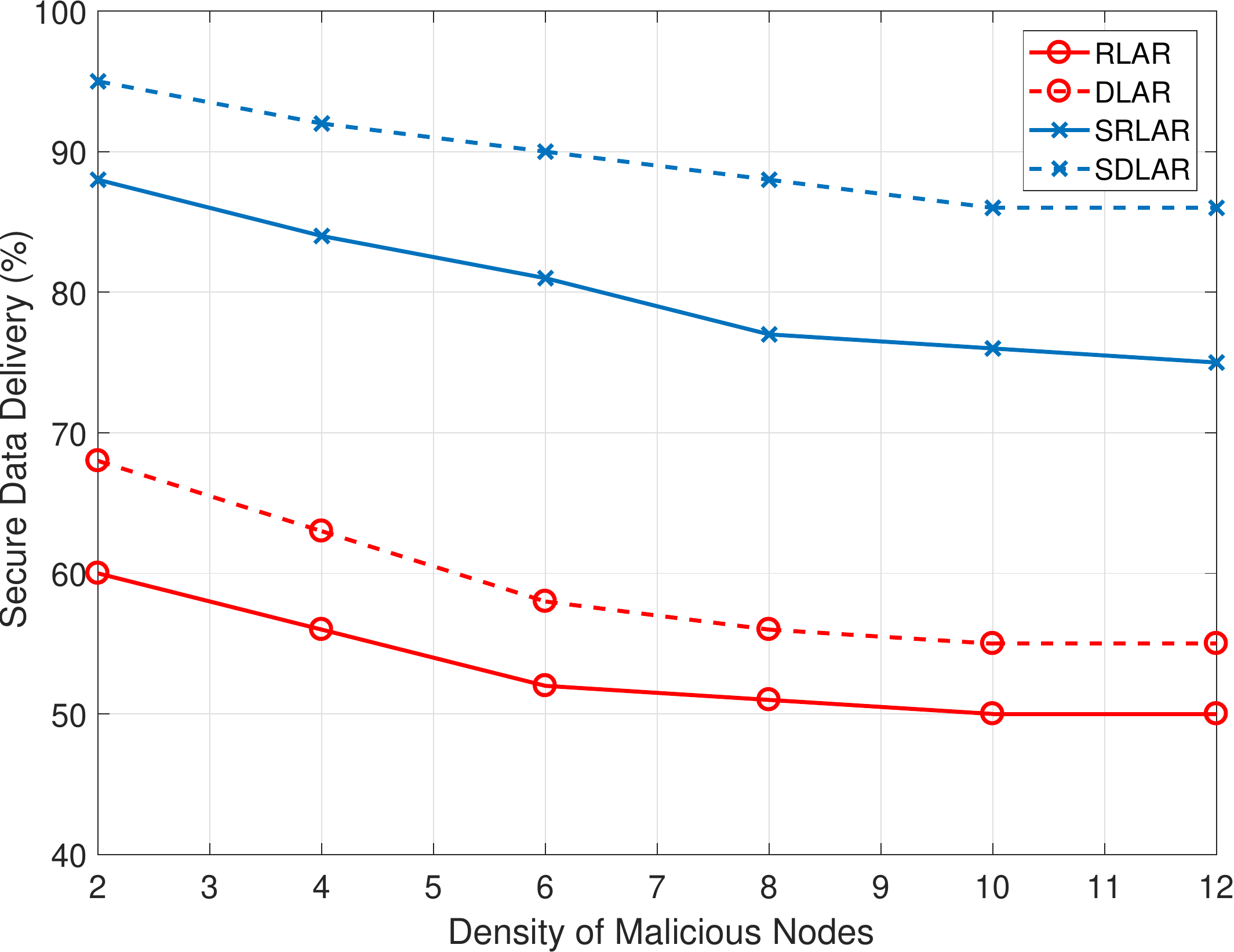}
\caption{Effect of malicious nodes on secure data delivery. 40 vehicular nodes.  Node speed is 5 units/sec.}
\label{fig:delivery_threat}
\end{figure}

\begin{figure}[!t]
\centering
\includegraphics[width=0.98\linewidth]{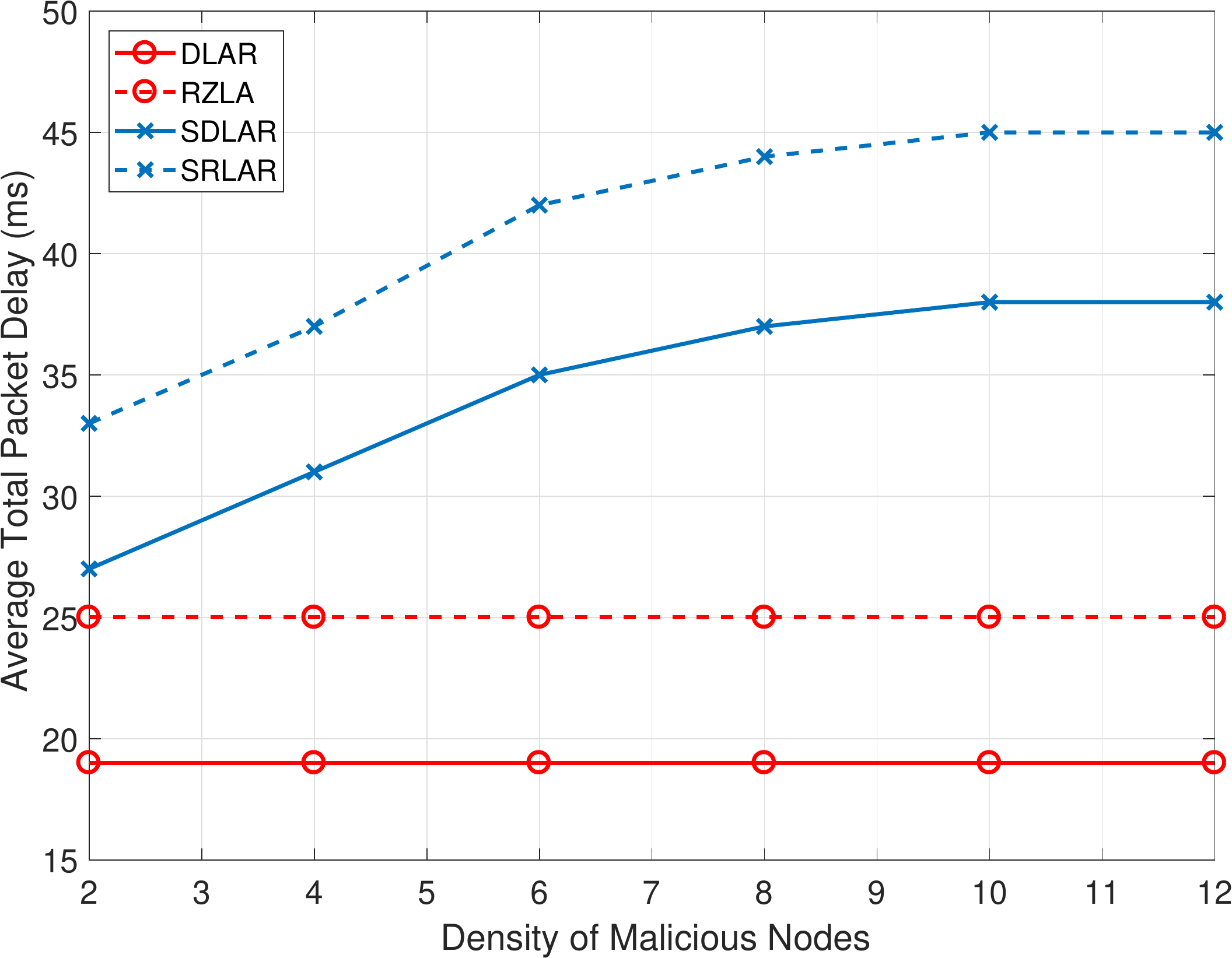}
\caption{Effect of malicious nodes on average total packet delays.  40 vehicular nodes.  Node speed is 5 units/sec.}
\label{fig:delay_threat}
\end{figure}

Fig.~\ref{fig:delivery_threat} and Fig.~\ref{fig:delay_threat} show that, with the secure routing protocol, a large number of malicious nodes has a negative effect on the VANET in reduced secure data delivery and increased average total packet delay.   The secure DLAR outperforms the secure RLAR, which makes it more suitable for the VANET with security threats.  
Fig.~\ref{fig:delay_threat} shows that the non-secure protocols have less packets delay and is irrelevant to the number of malicious nodes.  This is because these protocols do not have any security association phase which may lead to a catastrophe when facing security threats.

\begin{table}[!b]
\caption{Delivery (DLAR vs. SDLAR)} 
\centering 
\begin{tabular}{c c c c} 
\hline\hline 
\# of Mali-  & Delivery (DLAR) & Delivery (SDLAR) & Enhancement\\ 
cious Nodes &&& \\ [0.5ex] 
\hline 
2&	68&	95&	39.7\%\\
4&	63&	92&	46\%\\
6&	58&	90&	55.2\%\\
8&	56&	88&	57.1\%\\
10&	55&	86&	56.4\%\\
12&	55&	86&	56.4\%\\
\hline
 &  &  & Avg.=\text{~}51.8\% \\ [1ex] 
\end{tabular}
\label{table:delivery_DLAR_threat} 
\end{table}

\begin{table}[!b]
\caption{Delivery (RLAR vs. SRLAR)} 
\centering 
\begin{tabular}{c c c c} 
\hline\hline 
\# of Mali- & Delivery (RLAR) & Delivery (SRLAR) & Enhancement\\ 
cious Nodes &&& \\ [0.5ex] 
\hline 
2&	60&	88&	46.7\%\\
4&	56&	84&	50\%\\
6&	52&	81&	51.7\%\\
8&	51&	77&	55.8\%\\
10&	50&	76&	52\%\\
12&	50&	75&	50\%\\
\hline
 &  &  & Avg.=\text{~}51\% \\ [1ex] 
\end{tabular}
\label{table:delivery_RLAR_threat} 
\end{table}

\begin{table}[!b]
\caption{Delivery (SRLAR vs. SDLAR)} 
\centering 
\begin{tabular}{c c c c} 
\hline\hline 
\# of Mali- & Delivery (SRLAR) & Delivery (SDLAR) & Enhancement\\ 
cious Nodes &&& \\ [0.5ex] 
\hline 
2&	88&	95&	8\%\\
4&	84&	92&	9.5\%\\
6&	81&	90&	11.1\%\\
8&	77&	88&	14.3\%\\
10&	76&	86&	13.2\%\\
12&	75&	86&	14.7\%\\
\hline
 &  &  & Avg.=\text{~}11.8\% \\ [1ex] 
\end{tabular}
\label{table:delievery_SLAR_threat} 
\end{table}

\begin{table}[!b]
\caption{Delay (SRLAR vs. SDLAR))} 
\centering 
\begin{tabular}{c c c c} 
\hline\hline 
\# of Mali- & Delay (SRLAR) & Delay (SDLAR) & Enhancement\\ 
cious Nodes & [ms] & [ms] & \\ [0.5ex] 
\hline 
2&	33&	27&	18.2\%\\
4&	37&	31&	16.2\%\\
6&	42&	35&	16.7\%\\
8&	44&	37&	15.9\%\\
10&	45&	38&	15.6\%\\
12&	45&	38&	15.6\%\\
\hline
 &  &  & Avg.=\text{~}16.3\% \\ [1ex] 
\end{tabular}
\label{table:delay_SLAR_threat} 
\end{table}

The data are analyzed in the following tables to clarify the simulation results on routing performances with different numbers of malicious nodes in the VANET.  Table~\ref{table:delivery_DLAR_threat} and Table~\ref{table:delivery_RLAR_threat} show that the secure DLAR outperforms the non-secure DLAR in data delivery with an average 51.8\% enhancement and the secure RLAR outperforms the non-secure RLAR with an average 51\% enhancement.  
Table~\ref{table:delievery_SLAR_threat} and Table~\ref{table:delay_SLAR_threat} compare the two secure LAR protocols regarding secure data delivery and average total packet delay, respectively.  
 The tables show that, compared with the secure RLAR, the secure DLAR enhances the secure data delivery by an average 11.8\% and reduces the packet delay by an average 16.3\%.

\subsection{Node Speed Effect}

To show the effect of the node speed on the VANET performance metrics, i.e., secure data delivery and average packet delay, we simulate scenarios with the node speed $v = 5, 10, 15, 20, \ldots, 40$ units/sec.  The number of vehicular nodes in the VANET is 40, and the percentage of the malicious nodes is 10\%, i.e., 4 malicious nodes.

\begin{figure}[!t]
\centering
\includegraphics[width=0.98\linewidth]{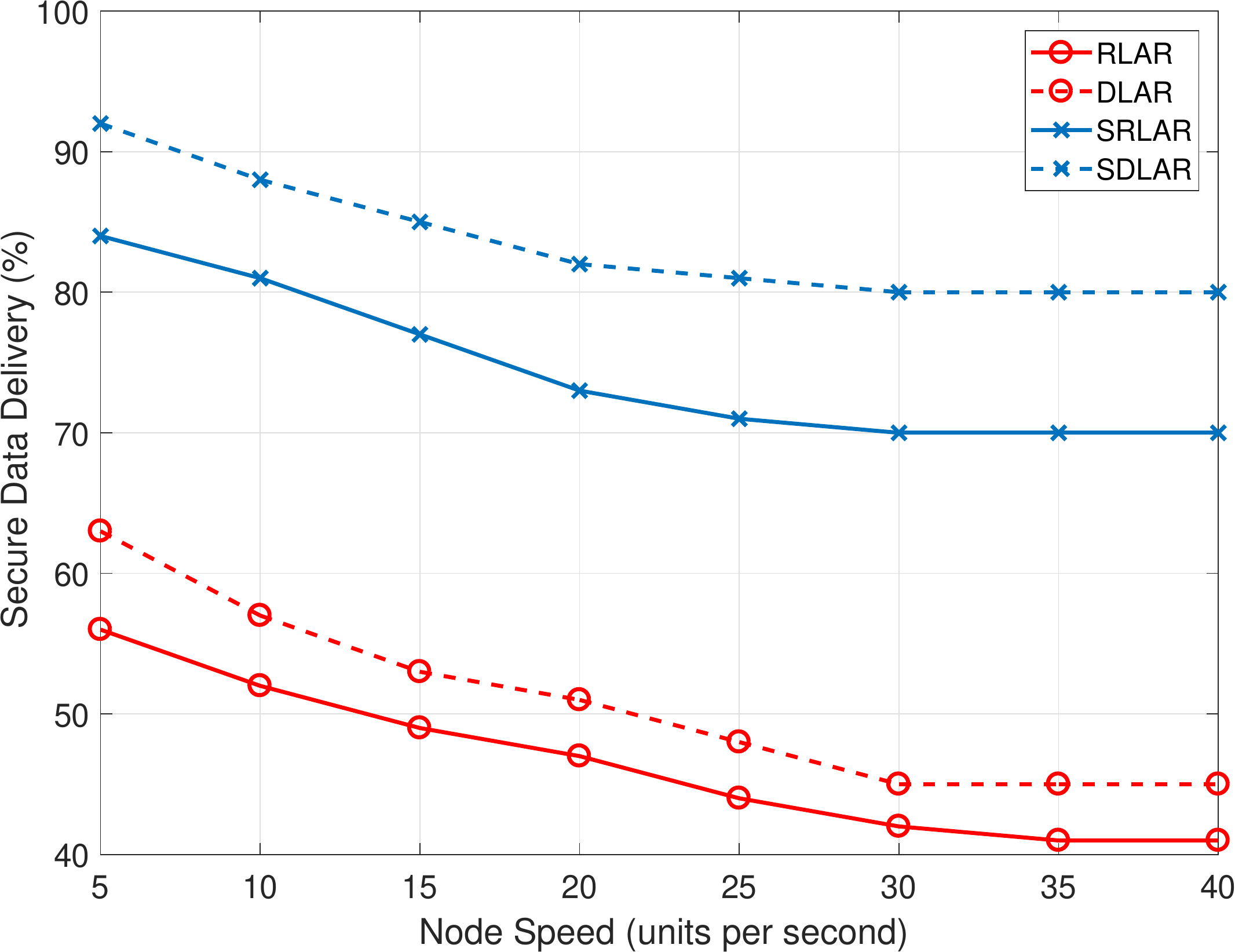}
\caption{Effect of node speed on secure data delivery.  40 vehicular nodes. 10\% malicious nodes.}
\label{fig:delivery_speed}
\end{figure}

\begin{figure}[!t]
\centering
\includegraphics[width=0.98\linewidth]{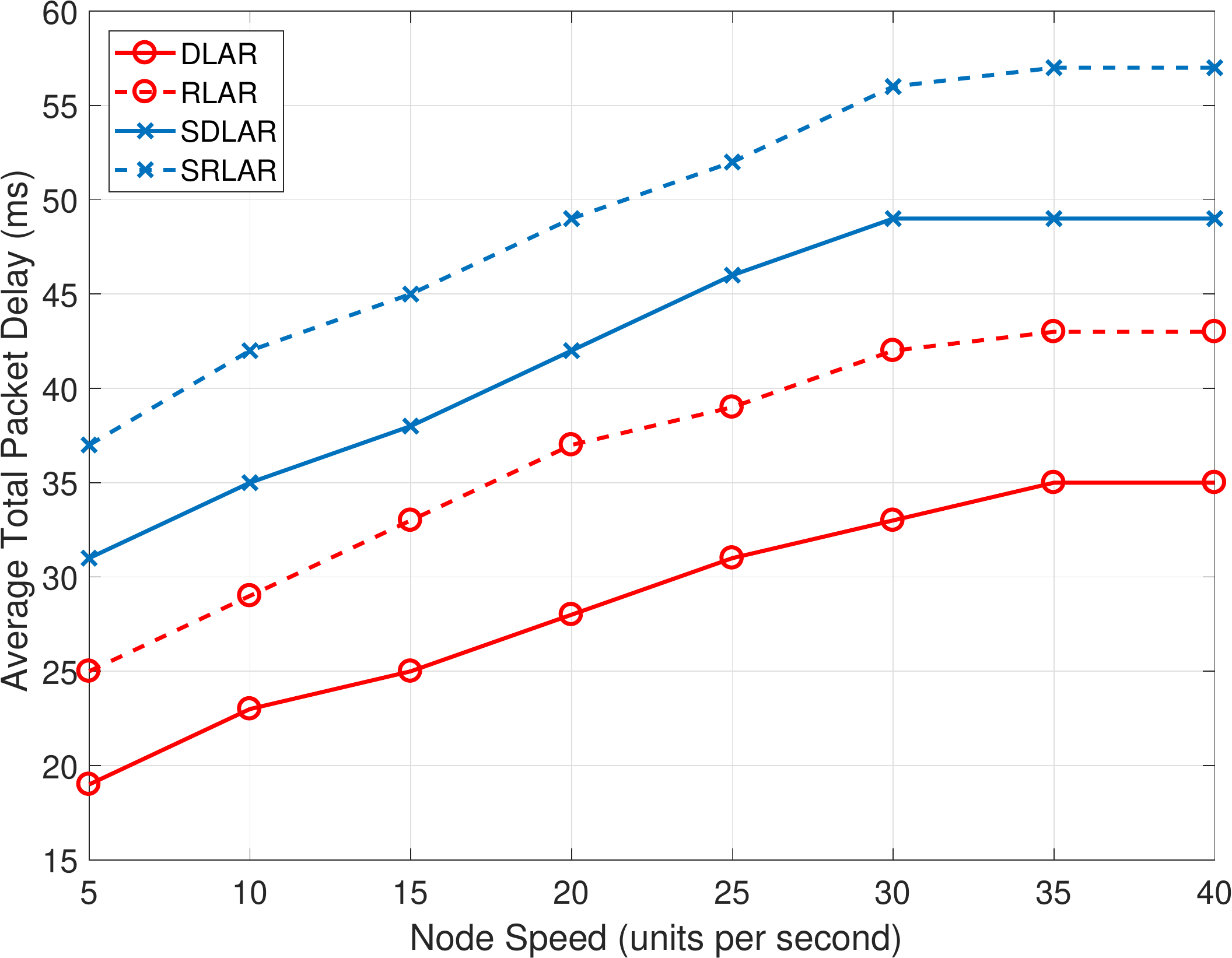}
\caption{Effect of node speed on average total packet delay.  40 vehicular nodes.  10\% malicious nodes.}
\label{fig:delay_speed}
\end{figure}

Fig.~\ref{fig:delivery_speed} and Fig.~\ref{fig:delay_speed} show that the higher speed at which the nodes move, the less secure data delivery and the larger average total packet delay there are in the VANET.  This is because the high mobility leads to the frequent route disconnection.  Therefore, the overhead is high of finding a stable route from the source to the destination.   The results show that the proposed secure LAR provides better data delivery with larger average total packet delay compared with the non-secure LAR.  And, the secure DLAR outperforms the secure RLAR in data delivery and packet delay.

\begin{table}[!b]
\caption{Delivery (DLAR vs. SDLAR)} 
\centering 
\begin{tabular}{c c c c} 
\hline\hline 
Node Speed & Delivery (DLAR) & Delivery (SDLAR) & Enhancement\\ 
(units/sec) &&& \\[0.5ex] 
\hline 
5&	63&	92&	46\%\\
10&	57&	88&	54.4\%\\
15&	53&	85&	60.4\%\\
20&	51&	82&	60.8\%\\
25&	48&	81&	68.8\%\\
30&	45&	80&	77.8\%\\
35&	45&	80&	77.8\%\\
40&	45&	80&	77.8\%\\
\hline
 &  &  & Avg.=\text{~}65.5\% \\ [1ex] 
\end{tabular}
\label{table:delivery_DLAR_speed} 
\end{table}

\begin{table}[!b]
\caption{Delivery (RLAR vs. SRLAR)} 
\centering 
\begin{tabular}{c c c c} 
\hline\hline 
Node Speed  & Delivery (RLAR) & Delivery (SRLAR) & Enhancement\\ 
(units/sec) &&& \\ [0.5ex] 
\hline 
5&	56&	84&	50\%\\
10&	52&	81&	55.8\%\\
15&	49&	77&	57.1\%\\
20&	47&	73&	55.3\%\\
25&	44&	71&	61.4\%\\
30&	42&	70&	66.7\%\\
35&	41&	70&	70.7\%\\
40&	41&	70&	70.7\%\\
\hline
 &  &  & Avg.=\text{~}60.9\% \\ [1ex] 
\end{tabular}
\label{table:delivery_RLAR_speed} 
\end{table}

\begin{table}[!b]
\caption{Delivery (SRLAR vs. SDLAR)} 
\centering 
\begin{tabular}{c c c c} 
\hline\hline 
Node Speed  & Delivery (SRLAR) & Delivery (SDLAR) & Enhancement\\ 
(units/sec) &&& \\[0.5ex] 
\hline 
5&	84&	92&	9.5\%\\
10&	81&	88&	8.6\%\\
15&	77&	85&	10.4\%\\
20&	73&	82&	12.3\%\\
25&	71&	81&	14.1\%\\
30&	70&	80&	14.3\%\\
35&	70&	80&	14.3\%\\
40&	70&	80&	14.3\%\\
\hline
 &  &  & Avg.=\text{~}12.2\% \\ [1ex] 
\end{tabular}
\label{table:delivery_SLAR_speed} 
\end{table}

\begin{table}[!b]
\caption{Delay (SRLAR vs. SDLAR))} 
\centering 
\begin{tabular}{c c c c} 
\hline\hline 
Node Speed  & Delay (SRLAR) & Delay (SDLAR) & Enhancement\\ 
(units/sec) & [ms] & [ms] & \\ [0.5ex] 
\hline 
5&	37&	31&	16.2\%\\
10&	42&	35&	16.7\%\\
15&	45&	38&	15.6\%\\
20&	49&	42&	14.3\%\\
25&	52&	46&	11.5\%\\
30&	56&	49&	12.5\%\\
35&	57&	49&	14\%\\
40&	57&	49&	14\%\\
\hline
 &  &  & Avg.=\text{~}14.4\% \\ [1ex] 
\end{tabular}
\label{table:delay_SLAR_speed} 
\end{table}

The performance of the VANET with different node speeds is clarified in the following tables.  
Table~\ref{table:delivery_DLAR_speed} and Table~\ref{table:delivery_RLAR_speed} show that the secure LAR routing protocols outperform the non-secure LAR routing protocols regarding secure data delivery with different node speeds.  The records in the tables show that the secure DLAR outperforms the non-secure DLAR with an average 65.5\% and the secure RLAR outperforms the non-secure RLAR with an average 60.9\% in data delivery.
Table~\ref{table:delivery_SLAR_speed} and Table~\ref{table:delay_SLAR_speed} compare the secure DLAR and the secure RLAR regarding secure data delivery and average total packet delay, respectively.  It is revealed that using the secure DLAR has an average 12.2\% in enhanced data delivery and an average 14.4\% in reduced packet delay.

\section{Conclusions}
\label{sec:conclusion}

Two secure location-aided routing (LAR) protocols are proposed for the VANET.  One routing protocol is based on the request zone and the other on the distance to the destination node.  The protocols use Diffie-Hellman key agreement protocol with short authentication strings to establish secure communication links between vehicular nodes through Wi-Fi Direct.  The VANET is therefore protected against security threats such as the MITMA.  Extensive simulations are performed through the .Net framework to accommodate the dynamic geographic routing features.  With different network densities, security threats and node speeds, simulation results show that the proposed secure LAR methods improve secure data delivery with a tradeoff in average total packet delay.  Of the two proposed secure LAR methods, the secure DLAR outperforms the secure RLAR regarding both data delivery and packet delay.

\bibliographystyle{IEEEtran}
\bibliography{IEEEabrv,cognitiveradio,bargain_paper}

\begin{thebibliography}{10}
\providecommand{\url}[1]{#1}
\csname url@samestyle\endcsname
\providecommand{\newblock}{\relax}
\providecommand{\bibinfo}[2]{#2}
\providecommand{\BIBentrySTDinterwordspacing}{\spaceskip=0pt\relax}
\providecommand{\BIBentryALTinterwordstretchfactor}{4}
\providecommand{\BIBentryALTinterwordspacing}{\spaceskip=\fontdimen2\font plus
\BIBentryALTinterwordstretchfactor\fontdimen3\font minus
  \fontdimen4\font\relax}
\providecommand{\BIBforeignlanguage}[2]{{%
\expandafter\ifx\csname l@#1\endcsname\relax
\typeout{** WARNING: IEEEtran.bst: No hyphenation pattern has been}%
\typeout{** loaded for the language `#1'. Using the pattern for}%
\typeout{** the default language instead.}%
\else
\language=\csname l@#1\endcsname
\fi
#2}}
\providecommand{\BIBdecl}{\relax}
\BIBdecl

\bibitem{5343061}
J.~Harri, F.~Filali, and C.~Bonnet, ``Mobility models for vehicular ad hoc
  networks: a survey and taxonomy,'' \emph{IEEE Communications Surveys
  Tutorials}, vol.~11, no.~4, pp. 19--41, Fourth 2009.

\bibitem{raw2015analytical}
R.~S. Raw, D.~Lobiyal, S.~Das, and S.~Kumar, ``Analytical evaluation of
  directional-location aided routing protocol for {VANET}s,'' \emph{Wireless
  Personal Communications}, vol.~82, no.~3, pp. 1877--1891, 2015.

\bibitem{6118326}
Y.~C. Chu and N.~F. Huang, ``An efficient traffic information forwarding
  solution for vehicle safety communications on highways,'' \emph{IEEE
  Transactions on Intelligent Transportation Systems}, vol.~13, no.~2, pp.
  631--643, Jun. 2012.

\bibitem{6697835}
Z.~Li and C.~T. Chigan, ``On joint privacy and reputation assurance for
  vehicular ad hoc networks,'' \emph{IEEE Transactions on Mobile Computing},
  vol.~13, no.~10, pp. 2334--2344, Oct. 2014.

\bibitem{5426524}
H.~Xu, X.~Wu, H.~R. Sadjadpour, and J.~J. Garcia-Luna-Aceves, ``A unified
  analysis of routing protocols in {MANET}s,'' \emph{IEEE Transactions on
  Communications}, vol.~58, no.~3, pp. 911--922, Mar. 2010.

\bibitem{6583337}
Z.~Wang, Y.~Chen, and C.~Li, ``{PSR}: A lightweight proactive source routing
  protocol for mobile ad hoc networks,'' \emph{IEEE Transactions on Vehicular
  Technology}, vol.~63, no.~2, pp. 859--868, Feb. 2014.

\bibitem{6708418}
J.~M. Chang, P.~C. Tsou, I.~Woungang, H.~C. Chao, and C.~F. Lai, ``Defending
  against collaborative attacks by malicious nodes in {MANET}s: A cooperative
  bait detection approach,'' \emph{IEEE Systems Journal}, vol.~9, no.~1, pp.
  65--75, Mar. 2015.

\bibitem{7546607}
A.~K. Sharma and M.~C. Trivedi, ``Performance comparison of {AODV}, {ZRP} and
  {AODVDR} routing protocols in {MANET},'' in \emph{International Conference on
  Computational Intelligence Communication Technology (CICT)}, Feb. 2016, pp.
  231--236.

\bibitem{7286854}
M.~H. Eiza, T.~Owens, Q.~Ni, and Q.~Shi, ``Situation-aware {QoS} routing
  algorithm for vehicular ad hoc networks,'' \emph{IEEE Transactions on
  Vehicular Technology}, vol.~64, no.~12, pp. 5520--5535, Dec. 2015.

\bibitem{6060927}
H.~Saleet, R.~Langar, K.~Naik, R.~Boutaba, A.~Nayak, and N.~Goel,
  ``Intersection-based geographical routing protocol for {VANET}s: A proposal
  and analysis,'' \emph{IEEE Transactions on Vehicular Technology}, vol.~60,
  no.~9, pp. 4560--4574, Nov. 2011.

\bibitem{7375000}
K.~Pandey, S.~K. Raina, and R.~S. Rao, ``Hop count analysis of location aided
  multihop routing protocols for {VANET}s,'' in \emph{International Conference
  on Signal Processing, Computing and Control (ISPCC)}, Sep. 2015, pp. 68--73.

\bibitem{6563165}
C.~Wu, S.~Ohzahata, and T.~Kato, ``Flexible, portable, and practicable solution
  for routing in {VANET}s: A fuzzy constraint q-learning approach,'' \emph{IEEE
  Transactions on Vehicular Technology}, vol.~62, no.~9, pp. 4251--4263, Nov.
  2013.

\bibitem{6684213}
D.~Tian, Y.~Wang, H.~Xia, and F.~Cai, ``Clustering multi-hop information
  dissemination method in vehicular ad hoc networks,'' \emph{IET Intelligent
  Transport Systems}, vol.~7, no.~4, pp. 464--472, Dec. 2013.

\bibitem{6544635}
N.~Alam and A.~G. Dempster, ``Cooperative positioning for vehicular networks:
  Facts and future,'' \emph{IEEE Transactions on Intelligent Transportation
  Systems}, vol.~14, no.~4, pp. 1708--1717, Dec. 2013.

\bibitem{7401382}
N.~I. Shuhaimi, Heriansyah, and T.~Juhana, ``Comparative performance evaluation
  of {DSRC} and {Wi-Fi} {D}irect in {VANET},'' in \emph{International
  Conference on Instrumentation, Communications, Information Technology, and
  Biomedical Engineering (ICICI-BME)}, Nov. 2015, pp. 298--303.

\bibitem{7579020}
W.~Shen, B.~Yin, X.~Cao, L.~X. Cai, and Y.~Cheng, ``Secure device-to-device
  communications over {WiFi} direct,'' \emph{IEEE Network}, vol.~30, no.~5, pp.
  4--9, Sept./Oct. 2016.

\bibitem{4537038}
A.~Tufail, M.~Fraser, A.~Hammad, K.~K. Hyung, and S.-W. Yoo, ``An empirical
  study to analyze the feasibility of {WIFI} for {VANET}s,'' in
  \emph{International Conference on Computer Supported Cooperative Work in
  Design}, Apr. 2008, pp. 553--558.

\bibitem{6472013}
S.~K. Dhurandher, M.~S. Obaidat, A.~Jaiswal, A.~Tiwari, and A.~Tyagi,
  ``Vehicular security through reputation and plausibility checks,'' \emph{IEEE
  Systems Journal}, vol.~8, no.~2, pp. 384--394, Jun. 2014.

\bibitem{6565569}
M.~Saleh and L.~Dong, ``Real-time scheduling with security enhancement for
  packet switched networks,'' \emph{IEEE Transactions on Network and Service
  Management}, vol.~10, no.~3, pp. 271--285, Sep. 2013.

\bibitem{7127003}
F.~Qu, Z.~Wu, F.~Y. Wang, and W.~Cho, ``A security and privacy review of
  {VANET}s,'' \emph{IEEE Transactions on Intelligent Transportation Systems},
  vol.~16, no.~6, pp. 2985--2996, Dec. 2015.

\bibitem{7754846}
R.~Mishra, A.~Singh, and R.~Kumar, ``{VANET} security: Issues, challenges and
  solutions,'' in \emph{International Conference on Electrical, Electronics,
  and Optimization Techniques (ICEEOT)}, Mar. 2016, pp. 1050--1055.

\bibitem{6166905}
J.~Toutouh, J.~Garcia-Nieto, and E.~Alba, ``Intelligent {OLSR} routing protocol
  optimization for {VANET}s,'' \emph{IEEE Transactions on Vehicular
  Technology}, vol.~61, no.~4, pp. 1884--1894, May 2012.

\bibitem{6512236}
J.~A. Martinez, D.~Vigueras, F.~J. Ros, and P.~M. Ruiz, ``Evaluation of the use
  of guard nodes for securing the routing in {VANET}s,'' \emph{Journal of
  Communications and Networks}, vol.~15, no.~2, pp. 122--131, Apr. 2013.

\bibitem{4640905}
V.~Pathak, D.~Yao, and L.~Iftode, ``Securing location aware services over
  {VANET} using geographical secure path routing,'' in \emph{IEEE International
  Conference on Vehicular Electronics and Safety}, Sep. 2008, pp. 346--353.

\bibitem{4349671}
C.~Harsch, A.~Festag, and P.~Papadimitratos, ``Secure position-based routing
  for {VANET}s,'' in \emph{IEEE 66th Vehicular Technology Conference}, Sep.
  2007, pp. 26--30.

\bibitem{7425245}
S.~Jiang, X.~Zhu, and L.~Wang, ``An efficient anonymous batch authentication
  scheme based on {HMAC} for {VANET}s,'' \emph{IEEE Transactions on Intelligent
  Transportation Systems}, vol.~17, no.~8, pp. 2193--2204, Aug. 2016.

\bibitem{5376147}
S.~Lu, L.~Li, K.~Y. Lam, and L.~Jia, ``{SAODV}: A {MANET} routing protocol that
  can withstand black hole attack,'' in \emph{International Conference on
  Computational Intelligence and Security}, vol.~2, Dec. 2009, pp. 421--425.

\bibitem{ko2000location}
Y.-B. Ko and N.~H. Vaidya, ``Location-aided routing ({LAR}) in mobile ad hoc
  networks,'' \emph{Wireless networks}, vol.~6, no.~4, pp. 307--321, 2000.

\bibitem{pass2003deniability}
R.~Pass, ``On deniability in the common reference string and random oracle
  model,'' in \emph{Annual International Cryptology Conference}.\hskip 1em plus
  0.5em minus 0.4em\relax Springer, 2003, pp. 316--337.

\end{thebibliography}

\end{document}